\documentclass[aps,prd,twocolumn,nofootinbib]{revtex4-1}
\usepackage{epsfig}
\usepackage[colorlinks,linkcolor=blue,anchorcolor=blue,citecolor=blue,urlcolor=blue,breaklinks=true]{hyperref}
\usepackage{graphicx}
\usepackage{amsmath,bm}
\usepackage{color}
\usepackage{overpic}
\usepackage{multirow}
\usepackage{ulem}
\usepackage{epstopdf}
\usepackage{epsfig}

\newcommand{\pd}{\partial}
\newcommand{\m}{\mathrm}

\newcommand{\n}{\nonumber}
\newcommand{\abs}[1]{\left|#1\right|}
\newcommand{\rb}[1]{\left(#1\right)}
\renewcommand{\sb}[1]{\left[#1\right]}

\begin{document}
\author{Long Du$^{1}$, Yong-Long Wang$^{2,3}$, Guo-Hua Liang$^{1}$, Guang-Zhen Kang$^{2}$, Xiao-Jun Liu$^{1}$}\email{Email: liuxiaojun@nju.edu.cn}
\author{Hong-Shi Zong$^{2,4,5}$}\email[]{Email: zonghs@nju.edu.cn}
\address{$^{1}$ Key Laboratory of Modern Acoustics, Department of Physics, Collaborative Innovation Center of Advanced Microstructures, Nanjing University, Nanjing 210093, China}
\address{$^{2}$ Department of Physics, Nanjing University, Nanjing 210093, China}
\address{$^{3}$ Department of Physics, School of Science, Linyi University, Linyi 276005, China}
\address{$^{4}$ Joint Center for Particle, Nuclear Physics and Cosmology, Nanjing 210093, China}
\address{$^{5}$ State Key Laboratory of Theoretical Physics, Institute of Theoretical Physics, CAS, Beijing 100190, China}

\title{Curvature-induced bound states and coherent electron transport on the surface of a truncated cone}
\begin{abstract}
We study the curvature-induced bound states and the coherent transport properties for a particle constrained to move on a truncated cone-like surface.
With longitudinal hard wall boundary condition, the probability densities and spectra energy shifts are calculated, and are found to be obviously affected by the surface curvature.
The bound-state energy levels and energy differences decrease as increasing the vertex angle or the ratio of axial length to bottom radius of the truncated cone.
In a two-dimensional (2D) GaAs substrate with this geometric structure, an estimation of the ground-state energy shift of ballistic transport electrons induced by the geometric potential (GP) is addressed, which shows that the fraction of the ground-state energy shift resulting from the surface curvature is unnegligible under some region of geometric parameters.
Furthermore, we model a truncated cone-like junction joining two cylinders with different radii, and investigate the effect of the GP on the transmission properties by numerically solving the open-boundary 2D Schr\"odinger equation with GP on the junction surface.
It is shown that the oscillatory behavior of the transmission coefficient as a function of the injection energy is more pronounced when steeper GP wells appear at the two ends of the junction.
Moreover, at specific injection energy, the transmission coefficient is oscillating with the ratio of the cylinder radii at incoming and outgoing sides.

\bigskip


%68.65.-k Low-dimensional, mesoscopic, nanoscale and other related systems: structure and nonelectronic properties
%03.65.Ge Solutions of wave equations: bound states
%02.40.-k Geometry, differential geometry, and topology
\end{abstract}
\maketitle

\section{Introduction}
The quantum dynamics for a particle constrained to move on a 2D curved surface is a traditional subject that has provoked controversies for decades \cite{pr 85 635,ap 63 586,pra 23 1982,pra 25 2893,pra 80 052109}.
It is another important application of Riemann geometry in modern physics besides Einstein's theory of general relativity \cite{epl 98 27001,I B Khriplovich}.
With the advent and development of nanostructure and quantum waveguide technology\cite{prl 93 090407,pla 269 148}, the geometric effects were observed experimentally in some nano-devices \cite{epl 98 27001,prl 104 150403}.
For the constrained systems with novel geometries, a great deal of theoretical works have been reported \cite{prb 45 14100,pra 68 014102,pla 372 6141,prb 79 033404,pra 81 014102,pra 58 77}.
Furthermore,in the presence of electromagnetic (EM) field with a proper choice of gauge, Giulio Ferrari and Giampaolo Cuoghi derived the surface Schr\"odinger equation (SSE) for a spinless charged particle constrained on a general curved surface without source current perpendicular to the thin-film surface.\cite{pra 80 052109,prl 100 230403}.
Quite recently, we deduced the surface Pauli equation and obtained the additive spin connection GP for a spin charged particle constrained on a curved surface with EM field\cite{pra 90 042117}.

Nanostructures with revolution surface play an important role in quantum devices. Several studies have discussed the effect of the geometric potential to the electronic states \cite{pra 68 014102, prb 61 13730, prb 72 035403, prb 81 165419}. Geometric actions were investigated on some special surfaces of revolution\cite{pra 58 77,prb 79 201401,ss 601 22,V Atanasova3}.
And the tunneling conductance of connected carbon nanotubes was studied in 1996\cite{prb 53 2044}.
The junction which joins two straight carbon tubes with different radii can be modeled as a 2D truncated cone-like surface.
In this paper, we briefly review the thin-layer quantization scheme, and re-derive the SSE on the revolution surface with generatrix function, $f(z)$, in cylindrical coordinates in Sec. \ref{sec_schrodinger} \cite{ps 72 105,pra 58 77}.
In Sec. \ref{sec_bound}, we solve the eigenfunctions of the SSE on a truncated cone analytically.
With the axial hard wall boundary condition, bound states and axial probability densities are presented.
The energy levels and energy differences as a function of the vertex angle and of the ratio of axial length to bottom radius are calculated.
Moreover, we give an estimation of ground-state energy shift for the confined electrons resulting from the GP in a frustum cone-like ballistic transport GaAs substrate.
In Sec. \ref{sec_coherent}, the effect of the GP induced by surface curvature on the coherent electron transport properties is addressed for truncated cone-like junctions joining two straight cylinders with different radii.
We study the effective GP and the corresponding transmission coefficient related to the geometric parameters of this structure.
In Sec. \ref{sec_conclusion}, conclusions are presented.

\section{Schr\"odinger equation on a revolution surface\label{sec_schrodinger}}
Let us consider a particle constrained to move on a 2D regular surface $S$ that is embedded in 3D space and can be parameterized as $\bm{r}\rb{x(q^1,q^2),y(q^1,q^2),z(q^1,q^2)}$ with $q^1$ and $q^2$ being the curvilinear coordinates over $S$.
The portion of the space in an immediate neighborhood of $S$ can be described by
\begin{eqnarray}\label{e_R_r}
\bm{R}(q^1,q^2,q^3)=\bm{r}(q^1,q^2)+q^3 \bm{n}(q^1,q^2),
\end{eqnarray}
where $\bm{n}$ is the unit vector normal to $S$ and $q^3$ is the coordinate indicating the distance from $S$.
The Schr\"{o}dinger equation in the curvilinear coordinates $(q^1,q^2,q^3)$ for a free particle attached to $S$ with normal confining potential reads
\begin{eqnarray} \label{e_se2}
&&\m{i}\hbar\frac{\partial}{\partial t} \psi(q^1,q^2,q^3,t)\n\\
&&\qquad=\frac{-\hbar^2}{2m}\sqrt{G}^{-1} \pd_i\left( \sqrt{G} G^{ij} \pd_j\psi(q^1,q^2,q^3,t)\right) \n\\
&&\qquad\qquad+ V_\lambda(q^3) \psi(q^1,q^2,q^3,t),
\end{eqnarray}
where $\pd_i\equiv\pd/\pd q^i$ with $i,j=1,2,3$, the cotravariant tensor $G^{ij}$ indicates the inverse of 3D metric $G_{ij}\equiv\pd_i\bm{R}\cdot\pd_j\bm{R}$, $G$ is the determinant of $G_{ij}$, and $V_\lambda(q^3)$ represents an ideal squeezing potential that is perpendicular to $S$ and satisfies
\begin{eqnarray}
\lim_{\lambda\to \infty}V_\lambda(q^3)=\left\{
\begin{array}{ccc}
0,\ q^3=0, \\
\infty,\ q^3\neq 0.	
\end{array}
\right.
\end{eqnarray}
The relations between $G_{ij}$ and the 2D reduced metric $g_{ab}$ ($a,b=1,2$) defined on $S$ are
\begin{eqnarray}\label{e_mG}
\left\{
\begin{array}{ll}
G_{ab} = & g_{ab}+\sb{\alpha g+(\alpha g)^T}_{ab} q^3\\ &+ (\alpha g \alpha^T)_{ab} (q^3)^2 , \\
G_{a 3} = & G_{3 b}=0 \ , \ G_{33}=1 ,
\end{array}
\right.
\end{eqnarray}
where $\alpha$ denotes the Weingarten curvature tensor of surface $S$ with its elements satisfying the Gauss-Weingarten equations \cite{E Weisstein,Chern}.

\begin{figure}
	\begin{overpic}[width=0.23\textwidth,tics=10]{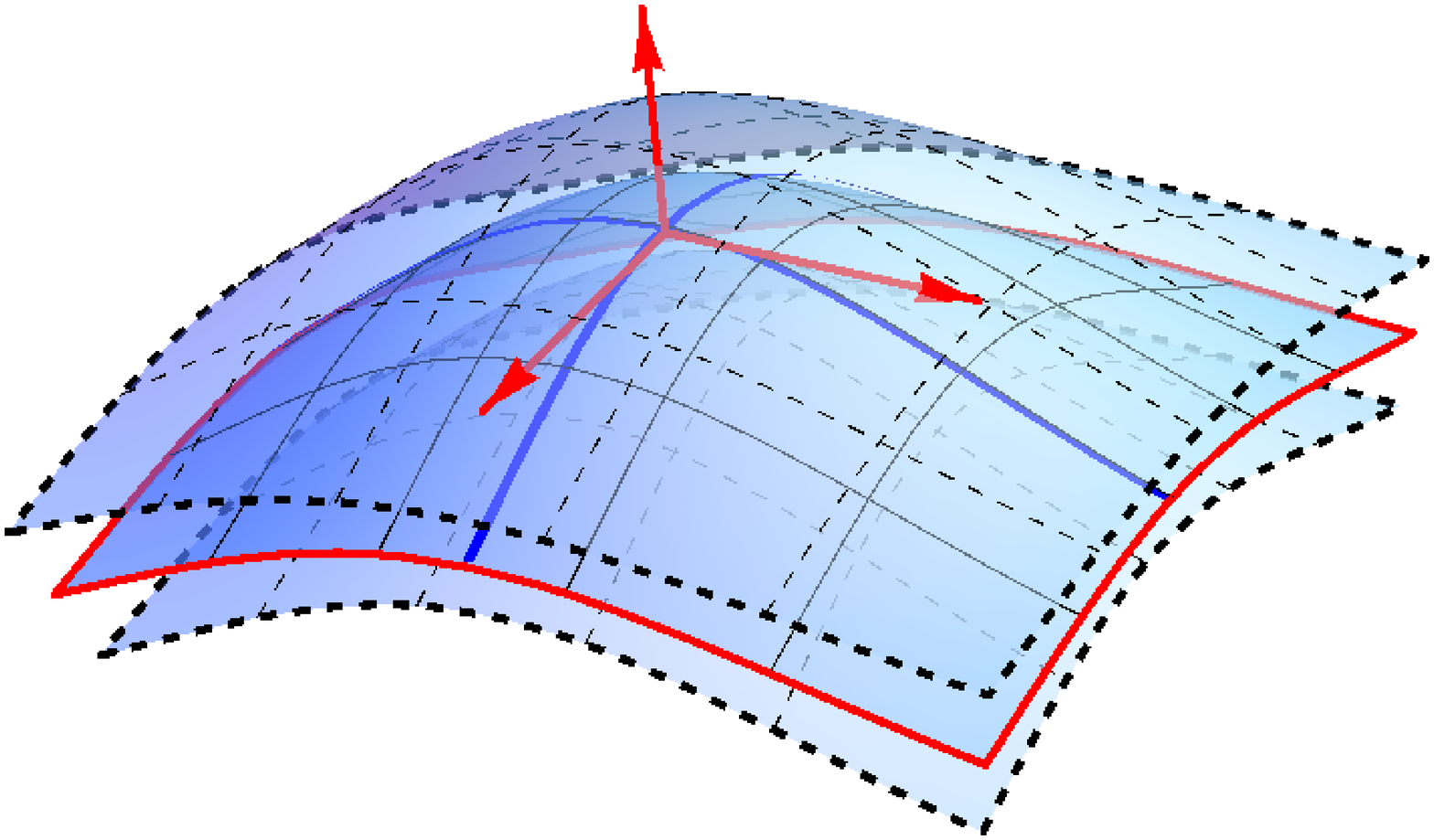}
		\put(45,67){$\bm{n}$}
		\put(23,47){\colorbox{white}{$\bm{e}_1$} }
		\put(70,50){\colorbox{white}{$\bm{e}_2$} }	
		\put(43,36){\colorbox{white}{$P$} }		
		\put(-3,20){$S$ }
		\put(-6,30){$S^\prime$ }
		\put(0,11){$S^{\prime\prime}$ }
		\put(28,15){$q^1$}
		\put(85,25){$q^2$}
		\put(45,-5){$(a)$}
	\end{overpic}
	\begin{overpic}[width=0.23\textwidth,tics=10]{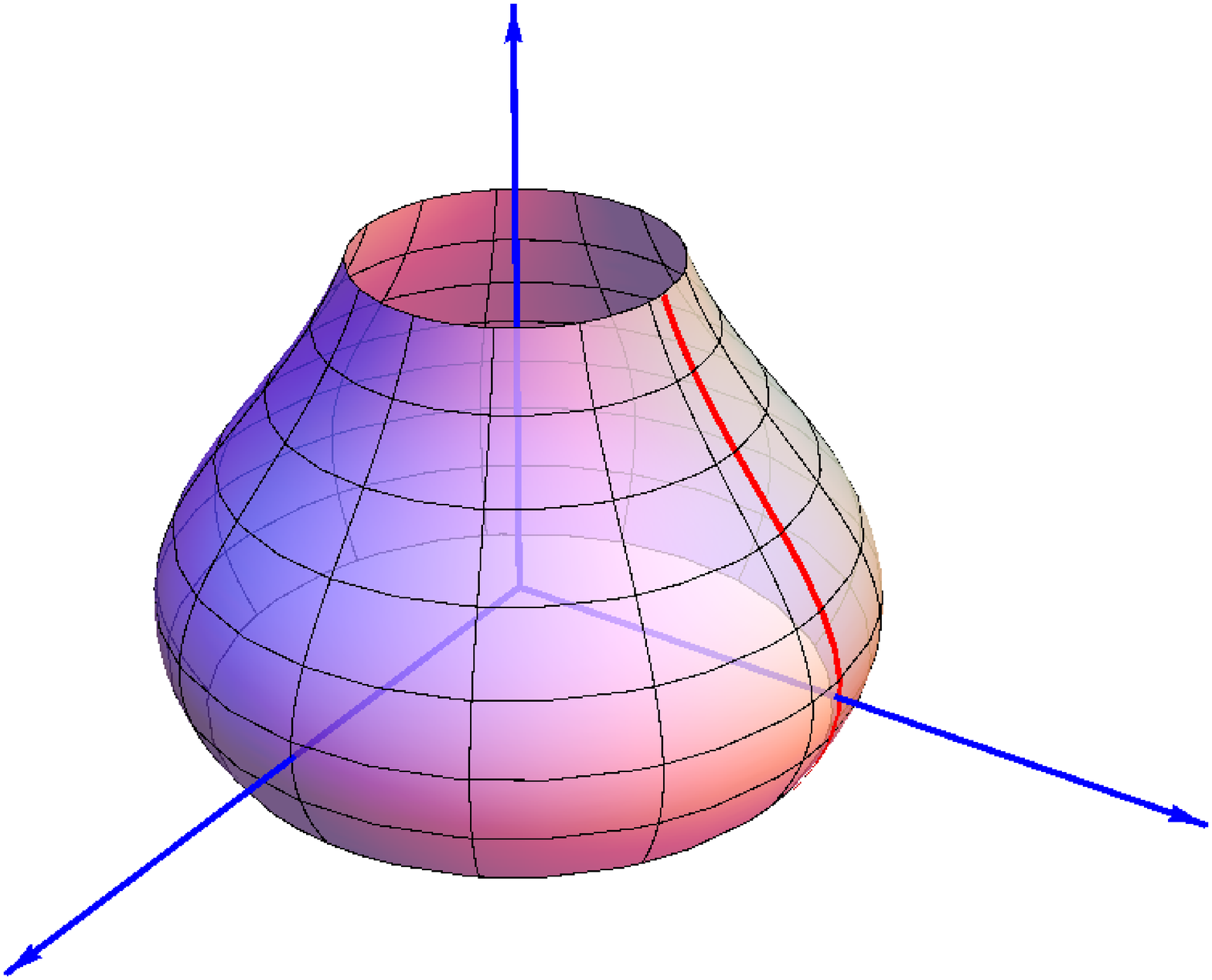}
		\put(0,16){$x$}
		\put(80,25){$y$}
		\put(38,72){$z$}	
		\put(60,50){$f(z)$}
		\put(60,50){\vector(-1,-2){5}}	
		\put(45,-5){$(b)$}
	\end{overpic}
	%	\begin{overpic}[width=0.32\textwidth,tics=10]{truncated_cone.png}
	%		\put(45,-5){$(c)$}
	%		\put(10,5){$x$}
	%		\put(90,25){$y$}
	%		\put(41,85){$z$}	
	%	\end{overpic}\\
	\caption{$(a)$ (color online) Sketch shows a two-dimensional curved surface $S$ with the curvilinear coordinates $(q^1,q^2)$ and two auxiliary surfaces $S',S''$.
		Normal vector $\bm{n}$ and $\bm{e}_1,\bm{e}_2$ form local orthogonal frame at any point $P$ that lies on the surface $S$.
		$(b)$ An arbitrary surface of revolution with generatrix $f(z)$.
	}\label{g_surfaces}
\end{figure}

Let $g\equiv\m{det}(g_{ab})$, the relation between $G$ and $g$ satisfies the following expression:
\begin{eqnarray}
G=\xi^2 g ,
\end{eqnarray}
where $\xi=1+\m{tr}(\alpha)q^3+\det(\alpha)(q^3)^2$.
After introducing a new wave function $\chi(q^1,q^2,q^3,t)=\chi_N(q^3,t)\chi_T(q^1,q^2,t)=\xi(q^1,q^2,q^3)^{1/2}\psi(q^1,q^2,q^3,t)$, where $\chi_N(q^3,t)$ and $\chi_T(q^1,q^2,t)$ represent the normal and tangent parts respectively,
and considering the limit $q^3\to 0$, Eq. (\ref{e_se2}) is separated into normal and surface components
\begin{eqnarray}
\frac{-\hbar^2}{2m} \pd_3^2 \chi_N + V_{\lambda}(q^3)\chi_N = \m{i}\hbar\frac{\partial}{\partial t}\chi_N, \label{e_se_norm_part}\\
\frac{-\hbar^2}{2m} \sb{ \frac{1}{\sqrt{g}}\pd_a \rb{\sqrt{g}g^{ab}\pd_b\chi_T } }+V_g\chi_T=\m{i}\hbar\frac{\partial}{\partial t} \chi_T, \label{e_se_tangent_part}
\end{eqnarray}
where the first term in Eq. (\ref{e_se_tangent_part}) stands for the kinetic term of the surface component \cite{ap 63 586,pra 58 77,pra 56 2592}, and $V_g=\frac{-\hbar^2}{2m}\left( \frac{1}{4}(\m{tr}(\alpha))^2-\det(\alpha)\right)$ is the well-known GP.
Since the confining potential , $V_\lambda(q^3)$, raises quantum excitation energy levels in the normal direction far beyond those in the tangential direction, Eq. (\ref{e_se_norm_part}) could be ignored \cite{pra 47 686}.

A surface of revolution $S_\m{rev}$ with generatrix $f(z)$ is cylindrically symmetric around the $z$ axis (Fig. \ref{g_surfaces}(b)).
Here, we assume that $f(z)$ is positive and analytic.
In the cylindrical coordinates, a point on $S_\m{rev}$ could be parametrized as $\bm{r}(\theta,z)=(f\cos\theta,f\sin\theta,z)$.
The local tangential derivative vectors at $\bm{r}(\theta,z)$ read
\begin{eqnarray}
\pd_\theta \bm{r}(\theta,z)=(-f\sin\theta,f\cos\theta,0),\\
\pd_z \bm{r}(\theta,z)=(f_{z}\cos\theta,f_{z}\sin\theta,1),
\end{eqnarray}
where $f_z$ represents $\frac{\pd}{\pd z} f$.
The covariant and contravariant reduced metric tensor on $S_\m{rev}$ are
\begin{eqnarray}
g_{ab}=\left(\begin{array}{ccc}
f^2 &0 \\
0 &1+f_{z}^2
\end{array}\right),\label{eq_g_ab} \\
g^{ab}=\left(\begin{array}{ccc}
f^{-2} & 0 \\
0 &(1+f_{z}^2)^{-1}
\end{array}\right). \label{eq_g^ab}
\end{eqnarray}
The unit normal vector $\bm{n}(\theta,z)=\bm{e}_\theta(\theta,z)\times\bm{e}_z(\theta,z)=\frac{1}{L(\bm{n})}(\cos\theta,\sin\theta,-f_{z})$ where $\bm{e}_\theta=\pd_\theta\bm{r}\left/\abs{\pd_\theta\bm{r}}\right.$, $\bm{e}_z=\pd_z\bm{r}\left/\abs{\pd_z\bm{r}}\right.$, and $L(\bm{n})=\sqrt{1+f_{z}^2}$.
The Weingarten curvature tensor in Eq. (\ref{e_mG}) reads
\begin{eqnarray}\label{e_Weingarten}
\alpha =\frac{1}{\sqrt{1+f_{z}^2}}
\left(\begin{array}{ccc}
\frac{1}{f} & 0\\
0 & \frac{-f_{zz}}{1+f_{z}^2}
\end{array}\right).
\end{eqnarray}

From Eq. (\ref{e_se_tangent_part}), we obtain the surface component Schr\"odinger equation as
\begin{eqnarray} \label{e_se_tangent_part2}
\frac{-\hbar^2}{2m}	\Bigg\{\frac{1}{f^2}\pd_\theta^2\chi_T&&+\frac{1}{1+f_{z}^2}\pd_z^2\chi_T +\frac{f_{z}(1+f_{z}^2-ff_{zz})}{f(1+f_{z}^2)^2}\n\\
&&\times\pd_z\chi_T \Bigg\} +U \chi_T =\m{i}\hbar\frac{\partial}{\partial t} \chi_T ,
\end{eqnarray}
where
\begin{eqnarray}\label{e_geo_revo}
	U = \frac{-\hbar^2}{2m} \frac{\rb{1+f_z^2+f f_{zz}^2}^2}{4f^2(1+f_z^2)^3},
\end{eqnarray}
is the GP on $S_\m{rev}$.
Replacing $\chi_T(\theta,z,t)$ by $\phi(\theta,z)\exp(-\m{i}\omega t/\hbar)$ and setting $\phi(\theta,z)=\Theta(\theta)Z(z)$, Eq. (\ref{e_se_tangent_part2}) is separated into two mutual independence second-order differential equations
\begin{eqnarray}
&&\pd_\theta^2\Theta+\eta^2\Theta=0, \label{e_se_theta} \\
&&\pd_z^2 Z+\frac{f_z(1+f_z^2-ff_{zz})}{f(1+f_z^2)}\pd_z Z+\Bigg[\frac{(1+f_z^2+f f_{zz})^2}{4f^2(1+f_z^2)^2} \n\\
&&\qquad +\frac{2m\omega}{\hbar^2}(1+f_z^2)-\frac{1+f_z^2}{f^2}\eta^2\Bigg]Z=0 \label{e_se_z} ,
\end{eqnarray}
where the eigenvalue $\eta^2$ of Eq. (\ref{e_se_theta}) is the separation constant that is independent of $\theta$ or $z$.

Before ending this section, we'd like to propose that it is unphysical for a particle to be perfectly constrained to a surface.
In the thin-layer quantization scheme, it shows that the confining potential is a good approximation for practical 2D nanostructures\cite{pra 58 77,sci 54 437}.
In the normal direction with the squeezing potential, quantum excitation energies are far beyond those in the tangential directions.
In this case, the ground states in the normal direction are preserved with fixed contribution to the total energy.
It means that any difference in the ground states comes from the surface part indicated by $\omega$ in  $\chi_T(\theta,z,t)$.
In the following, we will discuss the change of $\omega$ in the tangential part of eigenstates.

\section{Bound states and energy shifts on a truncated cone \label{sec_bound}}
\begin{figure}
	\begin{overpic}[width=0.48\textwidth,tics=10]{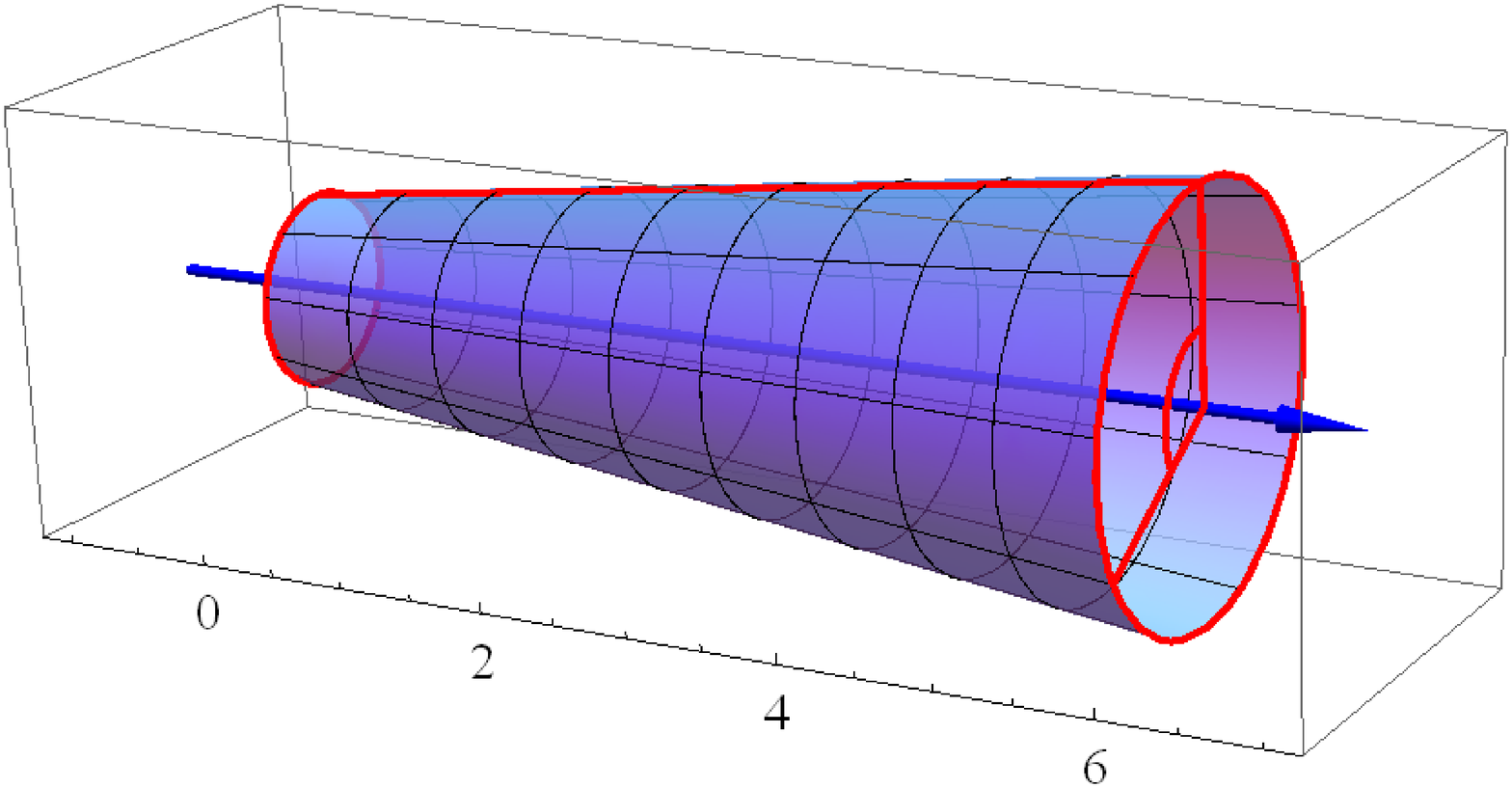}
		\put(20,40){\line(-6,1){8}}
		\put(15,46){\vector(0,-1){5}}
		\put(15,31){\vector(0,1){5}}		
		\put(14,38){$\rho$}
		\put(50,0){$(a)$}
		\put(91,27){{$z/\rho$}}
		\put(70,25){\colorbox{white}{$\theta$}}
	\end{overpic}\newline\newline\newline
	\begin{overpic}[width=0.48\textwidth,tics=10]{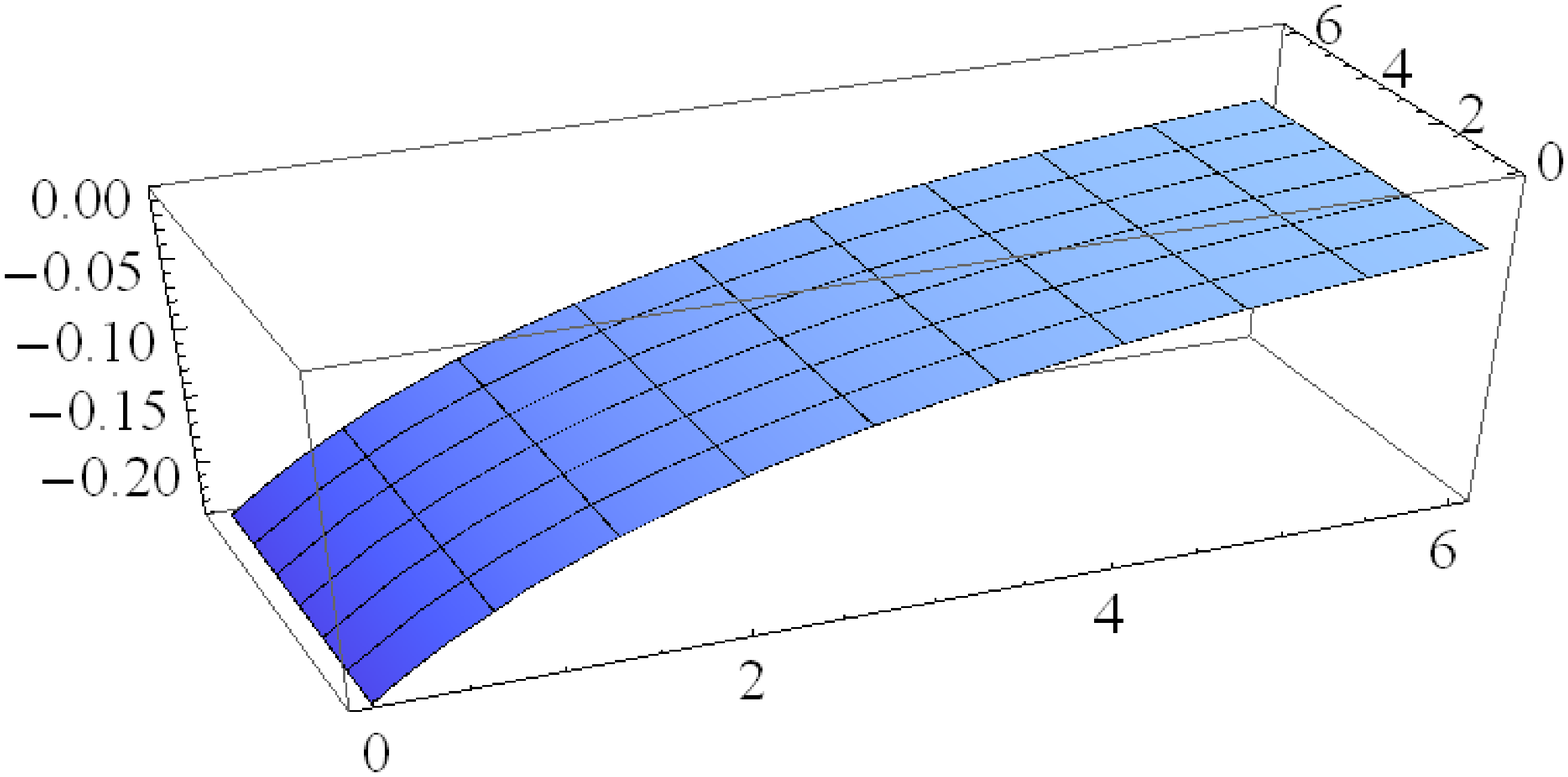}
		\put(50,0){$(b)$}		
		\put(60,7){$z/\rho$}
		\put(93,47){$\theta/\m{rad}$}
		\put(-6,15){\rotatebox{90}{$U\Big/\left(\frac{\hbar^2}{2m \rho^2}\right)$}}						
	\end{overpic}
	\caption{(color online) $(a)$ Surface of a truncated cone with the generatrix function $f(z)=\rho+\lambda z$ with $\rho$ indicating the radius of the smaller circular button.
		$(b)$ The GP of truncated cone depicted in $(a)$.
	}\label{g_truncated_cone}	
\end{figure}

The surface of a truncated cone (Fig. \ref{g_truncated_cone}$(a)$) can be parameterized as $\bm{r}(\theta,z)=\rb{\rb{\rho+\lambda z}\cos\theta,\rb{\rho+\lambda z}\sin\theta,z}$ , where $\theta$ and $z$ are the cylindrical coordinates over the truncated cone surface, $\rho$ denotes the radius of the smaller circular bottom, and $\lambda=\tan\beta$ with $\beta$ being the included angle between the generatrix and the $z$ axis.
Without loss of generality, we assume  $\lambda>0$.
From Eq. (\ref{e_se_z}), with $f_z=\lambda,f_{zz}=0$, we obtain
\begin{eqnarray}
\label{e_se_z2}&&\pd_z^2 Z+\frac{\lambda}{\rho+\lambda z}\pd_z Z+\Bigg[\frac{1}{(\rho+\lambda z)^2}\left(\frac{1}{4}-  \eta^2(1+\lambda^2)\right)\n\\
&&\qquad+2m\omega(1+\lambda^2)\Bigg]Z=0.
\end{eqnarray}
In consideration of the periodical and hard wall boundary conditions, $\Theta(\theta)$ and $Z(z)$ satisfy $\Theta(\theta+2\pi)=\Theta(\theta)$ and $Z(0)=Z(z_m)=0$ respectively, where $z_m$ represents the maximum of $z$.
For Eq. (\ref{e_se_theta}), the solutions are
\begin{eqnarray}
\Theta(\theta)=C \m{e}^{\m{i}\eta\theta},\quad(\eta=0,\pm 1,\pm 2,\cdots),
\end{eqnarray}
where $C$ is a nonzero complex constant.

By replacing $\rho+\lambda z$ by $y$, Eq. (\ref{e_se_z2}) can be rewritten as
\begin{eqnarray} \label{e_se_z3}
\pd_y^2 Z+ \frac{1}{y}\pd_y Z+\left[ \frac{1}{y^2}\left(\frac{1}{4\lambda^2}-\frac{\eta^2(1+\lambda^2)}{\lambda^2}\right)\right.\n\\
\left.\quad+\frac{2m\omega}{\hbar^2}\frac{1+\lambda^2}{\lambda^2} \right]Z=0 ,
\end{eqnarray}
which agrees with the form of Bessel equation with positive $2m\omega(1+\lambda^2)/(\hbar^2 \lambda^2)$.
The axial eigenstates for Eq. (\ref{e_se_z2}) are
\begin{eqnarray}\label{e_tr_cone_Z}
Z(z)= S_1 J_{\delta(\eta,\lambda)}\left(\frac{\sqrt{2m\omega(1+\lambda^2)}}{\hbar\lambda}(\rho+\lambda z)\right) \n\\
+S_2 Y_{\delta(\eta,\lambda)}\left(\frac{\sqrt{2m\omega(1+\lambda^2)}}{\hbar\lambda}(\rho+\lambda z)\right) ,
\end{eqnarray}
where $\delta(\eta,\lambda)=\lambda^{-1}\sqrt{\eta^2(1+\lambda^2)-1/4}$.
$J_{\delta(\eta,\lambda)}$ and $Y_{\delta(\eta,\lambda)}$ represent the Bessel and Neumann function of $\delta(\eta,\lambda)$ order respectively.
Coefficients $S_1$ and $S_2$ adjust the proportion of  these two kinds of Bessel function to meet the Dirichlet boundary condition on $z$. With the condition, the ratio of $-S_1$ to $S_2$ is
$\frac{Y_\delta\left[c\sqrt{1+\lambda^2}\lambda^{-1}\right]}{J_\delta\left[c\sqrt{1+\lambda^2}\lambda^{-1}\right]}$
, and the corresponding surface component energy $\omega$ is determined by the following equation
\begin{eqnarray}\label{e_ratio_det}
Y_\delta\left(c\sqrt{\frac{1+\lambda^2}{\lambda^2}}\right) J_\delta\left(c\sqrt{\frac{1+\lambda^2}{\lambda^2}}(1+\frac{z_m}{\rho}\lambda)\right)- \n\\
J_\delta\left(c\sqrt{\frac{1+\lambda^2}{\lambda^2}}\right) Y_\delta\left(c\sqrt{\frac{1+\lambda^2}{\lambda^2}}(1+\frac{z_m}{\rho}\lambda)\right)
=0,
\end{eqnarray}
where $c=\rho\sqrt{2m\omega}/\hbar$.
Obviously, Eq. (\ref{e_ratio_det}) is an algebraic equation of $c$ with multi-solutions which are denoted by $c_n$ sorted in ascending order.
Therefore, the eigenfunction Eq. (\ref{e_tr_cone_Z}) can be rewritten as
\begin{eqnarray}
Z_n(z)=A\Bigg\{J_\delta\left[c_n\sqrt{\frac{1+\lambda^2}{\lambda^2}}\bigg(1+\lambda\frac{z}{\rho}\bigg)\right] \n\\
-\frac{J_\delta\left(c_n \sqrt{\frac{1+\lambda^2}{\lambda^2}}\right)}{Y_\delta\left(c_n \sqrt{\frac{1+\lambda^2}{\lambda^2}}\right)} Y_\delta\left[c_n\sqrt{\frac{1+\lambda^2}{\lambda^2}}\bigg(1+\lambda\frac{z}{\rho}\bigg)\right] \Bigg\},
\end{eqnarray}
where $A$ represents the normalized coefficient.
Furthermore, $\omega_n=\frac{\hbar^2 c_n^2}{\rho^2 2m}$ are the corresponding surface component energy.

\begin{figure}
	\begin{overpic}[width=0.48\textwidth,tics=10]{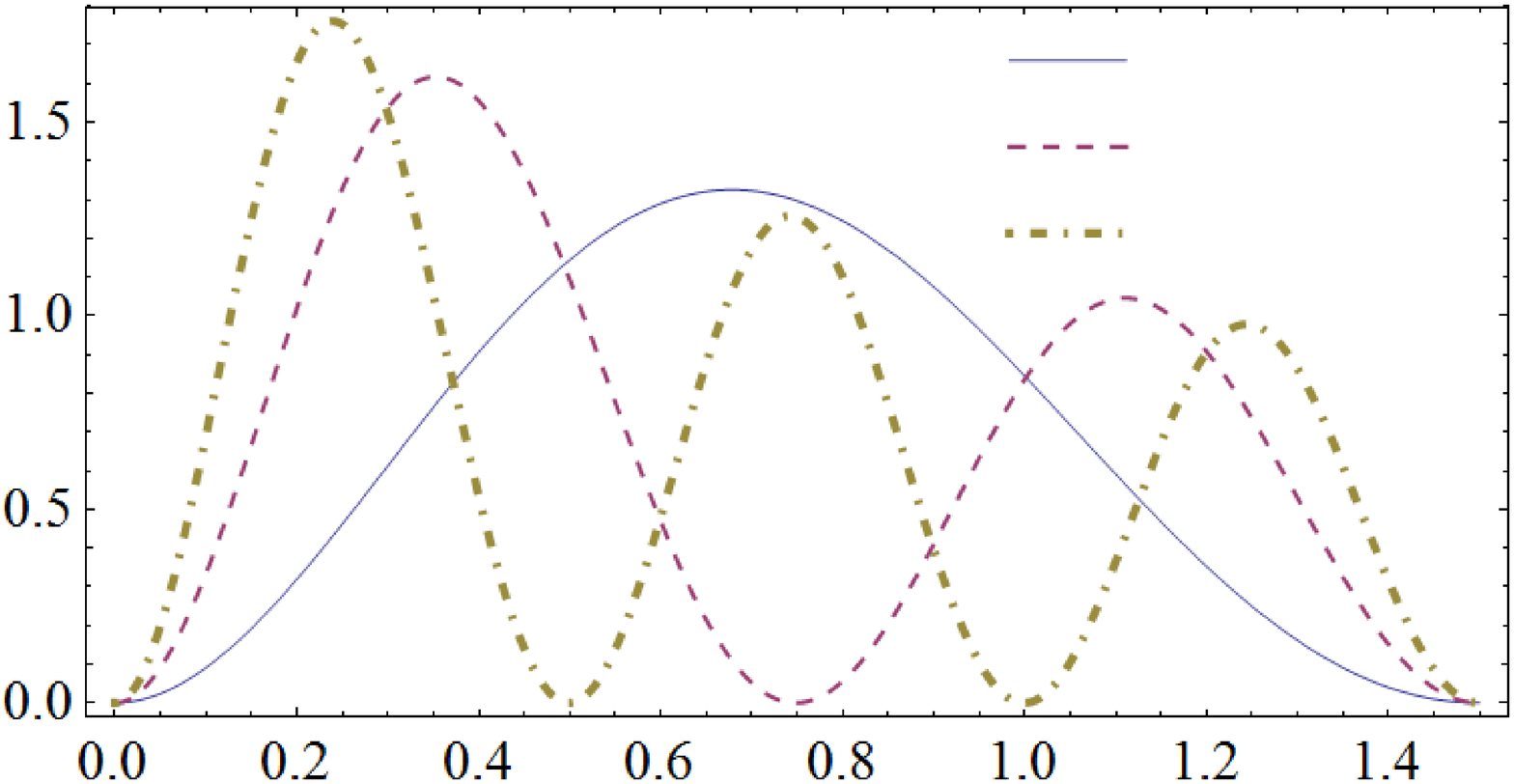}
		\put(37,48){{${z_m=1.5\rho}$}}
		\put(98,-3){$z/\rho$}
		\put(-5,25){\rotatebox{90}{PDs}}
		\put(60,35){\small{\rotatebox{90}{$\frac{\rho}{\hbar}\sqrt{2m\omega}$}}}
		\put(76,47){\small{$1.451$}}
		\put(76,41){\small{$2.946$}}
		\put(76,36){\small{$4.432$}}
		\put(50,-3){$(a)$}
	\end{overpic}\newline\newline\newline
	\begin{overpic}[width=0.48\textwidth,tics=10]{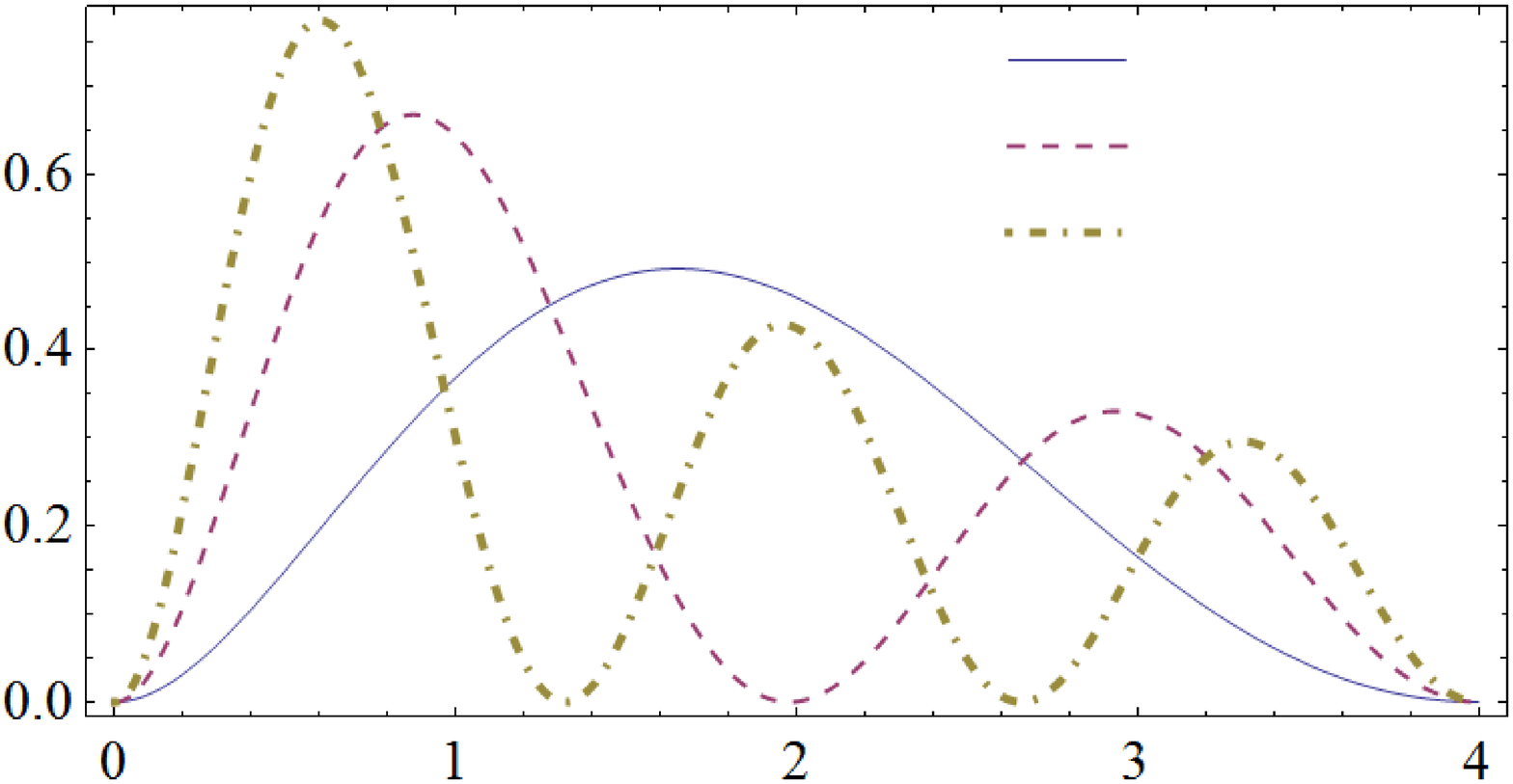}
		\put(37,48){{${z_m=4\rho}$}}
		\put(98,-3){$z/\rho$}
		\put(-5,25){\rotatebox{90}{PDs}}
		\put(60,35){\small{\rotatebox{90}{$\frac{\rho}{\hbar}\sqrt{2m\omega}$}}}
		\put(76,47){\small{$0.5233$}}
		\put(76,41){\small{$1.091$}}
		\put(76,36){\small{$1.652$}}		
		\put(50,-3){$(b)$}
	\end{overpic}		
	\caption{\label{g_pd1} (color online) Probability densities of the normalized axial eigenfunctions,  $Z_0(z),Z_1(z),Z_2(z)$, with $\eta=0,\lambda=1$. $(a)$ $z_m=1.5\rho$. $(b)$ $z_m=4\rho$}
\end{figure}

Tab. \ref{t_eta=0} classifies different values of $z_m/\rho$ of the truncated cone with $\eta^2=0$, $\lambda=1$.
For each fixed $z_m/\rho$, the three lowest eigenvalues of $\rho\sqrt{2m\omega}/\hbar$ are given. $\eta^2=0$ means $\Theta(\theta)$ is a constant and the angular momentum is zero.
The probability densities (PDs) depending on $z$ with $z_m/\rho=1.5$ are depicted in Fig. \ref{g_pd1}(a).
The curves which have one, two and three peaks respectively, denote the normalized PDs with respect to $\rho\sqrt{2m\omega}/\hbar=1.451,2.946$ and $4.432$.
Similarly, PDs with $z_m/\rho=4$ are illustrated in Fig. \ref{g_pd1}(b).
Graphics of PDs elucidate that the uneven GP affects the distribution of particles.
\begin{table}
	\begin{tabular}{ccc}
		\hline
		$z_m \rho^{-1}$ & $\rho\sqrt{2m\omega}/\hbar$ & $Z(z)$\\
		\hline
		\multirow{6}{*}{$1.5$} & \multirow{2}{*}{$1.451$} & $0.2905 \Big(J_{\frac{i}{2}}\left(\frac{2.052 (z+\rho )}{\rho }\right)$ \\
		& &$-(0.2323+0.7231 i)Y_{\frac{i}{2}}\left(\frac{2.052 (z+\rho )}{\rho }\right)\Big)$\\
		
		& \multirow{2}{*}{$2.946$} & $0.2957 \Big(J_{\frac{i}{2}}\left(\frac{4.166 (z+\rho )}{\rho }\right)$\\
		& & $-(0.2243+1.462 i) Y_{\frac{i}{2}}\left(\frac{4.166 (z+\rho )}{\rho }\right)\Big)$ \\
		
		& \multirow{2}{*}{$4.432$} & $0.1877 \Big(J_{\frac{i}{2}}\left(\frac{6.268 (z+\rho )}{\rho }\right)$\\
		& &$+(0.3795-0.8787 i) Y_{\frac{i}{2}}\left(\frac{6.268 (z+\rho )}{\rho }\right)\Big)$ \\
		\hline
		\multirow{6}{*}{$4$} & \multirow{2}{*}{$0.5233$} & $0.8447 \Big(J_{\frac{i}{2}}\left(\frac{0.74 (z+\rho )}{\rho }\right)$\\
		& & $+(0.3293-1.374 i) Y_{\frac{i}{2}}\left(\frac{0.74 (z+\rho )}{\rho }\right)\Big)$\\
		
		& \multirow{2}{*}{$1.091$} & $0.5364 \Big(J_{\frac{i}{2}}\left(\frac{1.543 (z+\rho )}{\rho }\right)$\\
		& & $-(0.4333+1.057 i) Y_{\frac{i}{2}}\left(\frac{1.543 (z+\rho )}{\rho }\right)\Big)$\\
		
		& \multirow{2}{*}{$1.652$} & $0.3495 \Big(J_{\frac{i}{2}}\left(\frac{2.337 (z+\rho )}{\rho }\right)$\\
		& &$-(0.06785+0.6611 i) Y_{\frac{i}{2}}\left(\frac{2.337 (z+\rho )}{\rho }\right)\Big)$ \\
		\hline	
	\end{tabular}
	\caption{\label{t_eta=0} The solutions of Eq. (\ref{e_se_z3}) with $\eta^2=0,\lambda=1$. The lowest three energy levels and normalised wavefunctions of $Z$ component are enumerated.}	
\end{table}

\begin{figure}[htb]{
		\begin{overpic}[width=0.22\textwidth,tics=10]{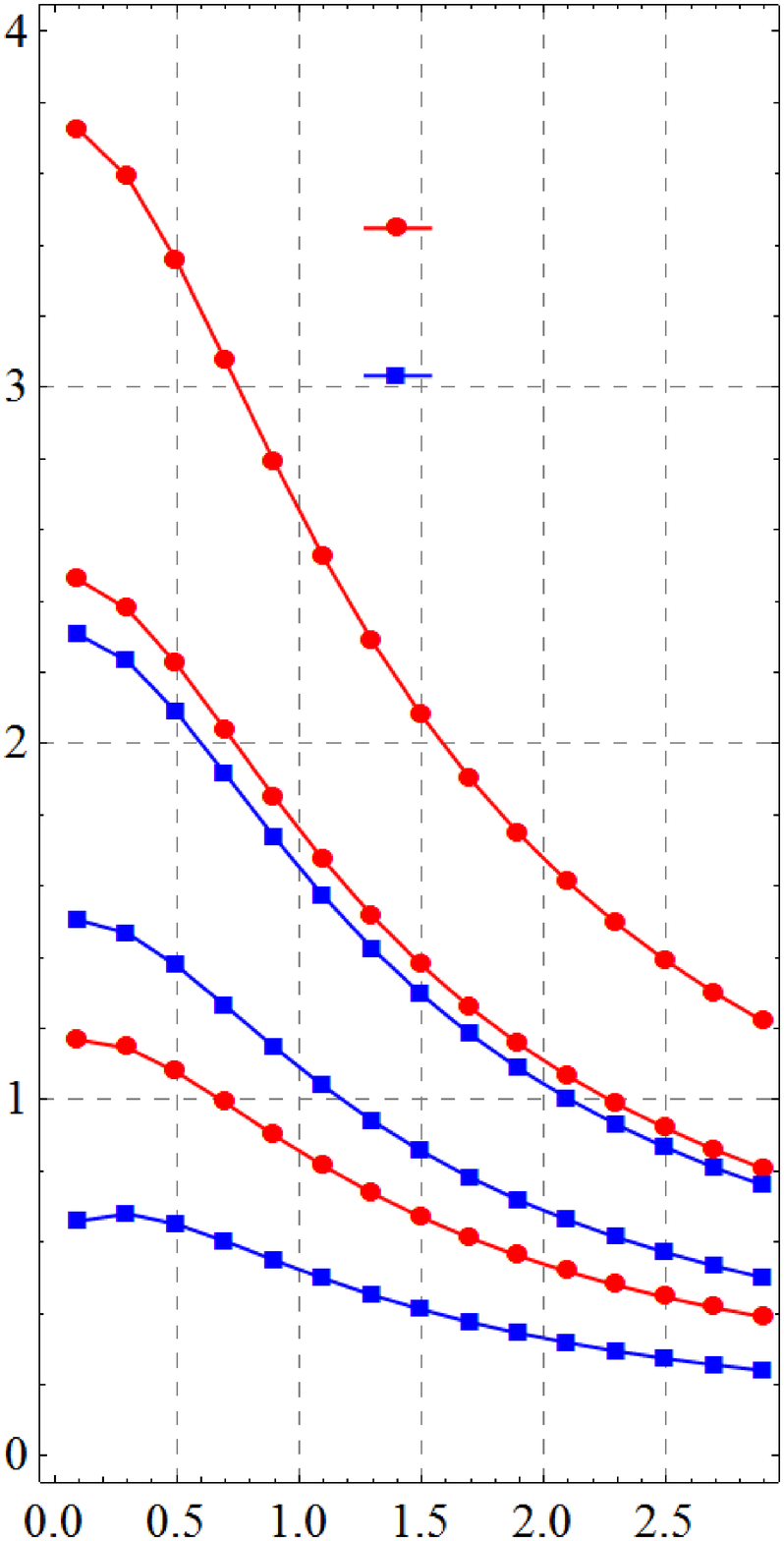}
			\put(20,-5){$(a)$}
			\put(18,92){\large{${\eta^2=0}$}}
			\put(-7,40){\rotatebox{90}{$\rho\sqrt{2m\omega}/\hbar$}}
			\put(45,-2){$\lambda$}
			\put(30,85){$\frac{z_m}{\rho}=2.5$}
			\put(30,75){$\frac{z_m}{\rho}=4$}			
		\end{overpic}
		\begin{overpic}[width=0.232\textwidth,tics=10]{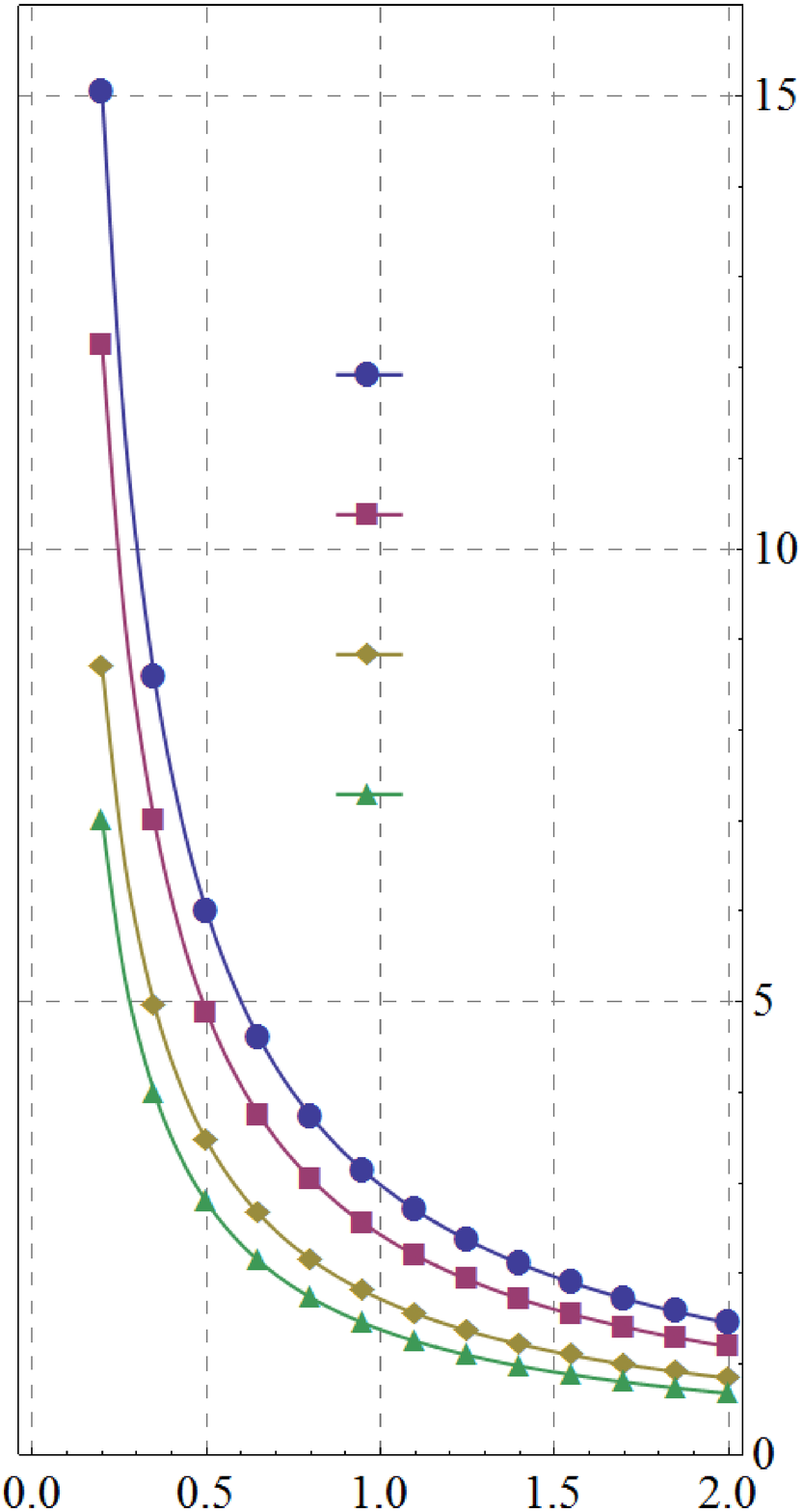}
			\put(20,-5){$(b)$}
			\put(20,92){\large{${\eta^2=0}$}}
			%		\put(54,65){\rotatebox{270}{$\rho\sqrt{2m\omega}/\hbar$}}
			\put(50,-2){$z_m/\rho$}
			\put(33,74){$\lambda=0.3$}
			\put(33,65){$\lambda=0.8$}			
			\put(33,56){$\lambda=1.5$}
			\put(33,47){$\lambda=2.0$}				
		\end{overpic}
	}
	\caption{\label{g_ES}(color online)
		$(a)$ The three lowest eigenvalues of $\rho\sqrt{2m\omega}/\hbar$ as a function of $\lambda$ with $z_m/\rho=2.5$ (red) and $z_m/\rho=4.0$ (blue) respectively.
		$(b)$ The lowest eigenvalue of $\rho\sqrt{2m\omega}/\hbar$ as a function of $z_m/\rho$ for $\lambda=0.3,0.8,1.5,2.0$ respectively.
	}
\end{figure}

The energy levels of bound states on the truncated cone depend on its geometric dimensions which are described parameterized by $\rho$, $\lambda$ and $z_m$.
Functional dependence between the values of the three lowest energy levels and slope of generatrix with different $z_m$ in the case of $\eta^2=0$ is illustrated in Fig. \ref{g_ES}$(a)$, which indicates that the
energy levels and energy differences monotonously decrease with increase of $\lambda$.
The relations between the energy shift of the ground states and the height of the truncated cone, $z_m$, with $\eta=0$ are shown in Fig. \ref{g_ES}$(b)$.
When $\lambda$ is fixed, increasing height will lower the ground-state energy levels.

When $\rho$ goes to zero, Eq. (\ref{e_se_z2}) describes the $z$ component wave function on a cone\cite{jmp 53 122106}.
With the hard wall boundary condition in $z$ direction, the Neumann function in Eq. (\ref{e_tr_cone_Z}) should be omitted because of its singularity at $z=0$.
With finite $\rho$, when $\lambda$ goes to zero Eq. (\ref{e_se_z2}) will become describing $z$ component wavefunction on the surface of a cylinder.

\begin{figure}[htb]
	\begin{overpic}[width=0.46\textwidth,tics=10]{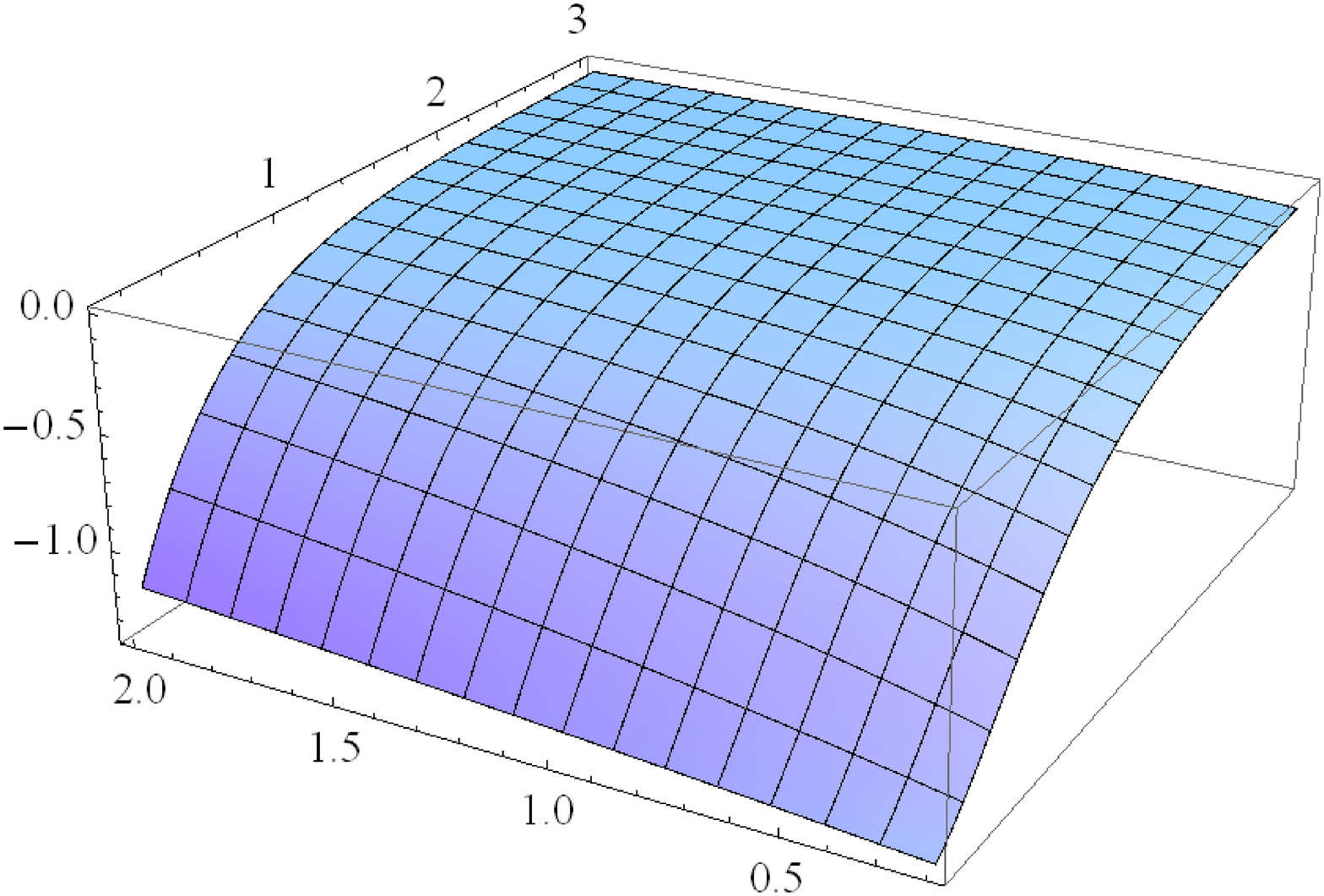}
		\put(23,60){$\lambda$}
		\put(31,2){$z_m/\rho$}
		\put(-5,18){\rotatebox{90}{$\langle\psi_0|U|\psi_0\rangle$/meV}}
		\put(45,-3){$(a)$}
	\end{overpic}\newline\newline\newline
	\begin{overpic}[width=0.46\textwidth,tics=10]{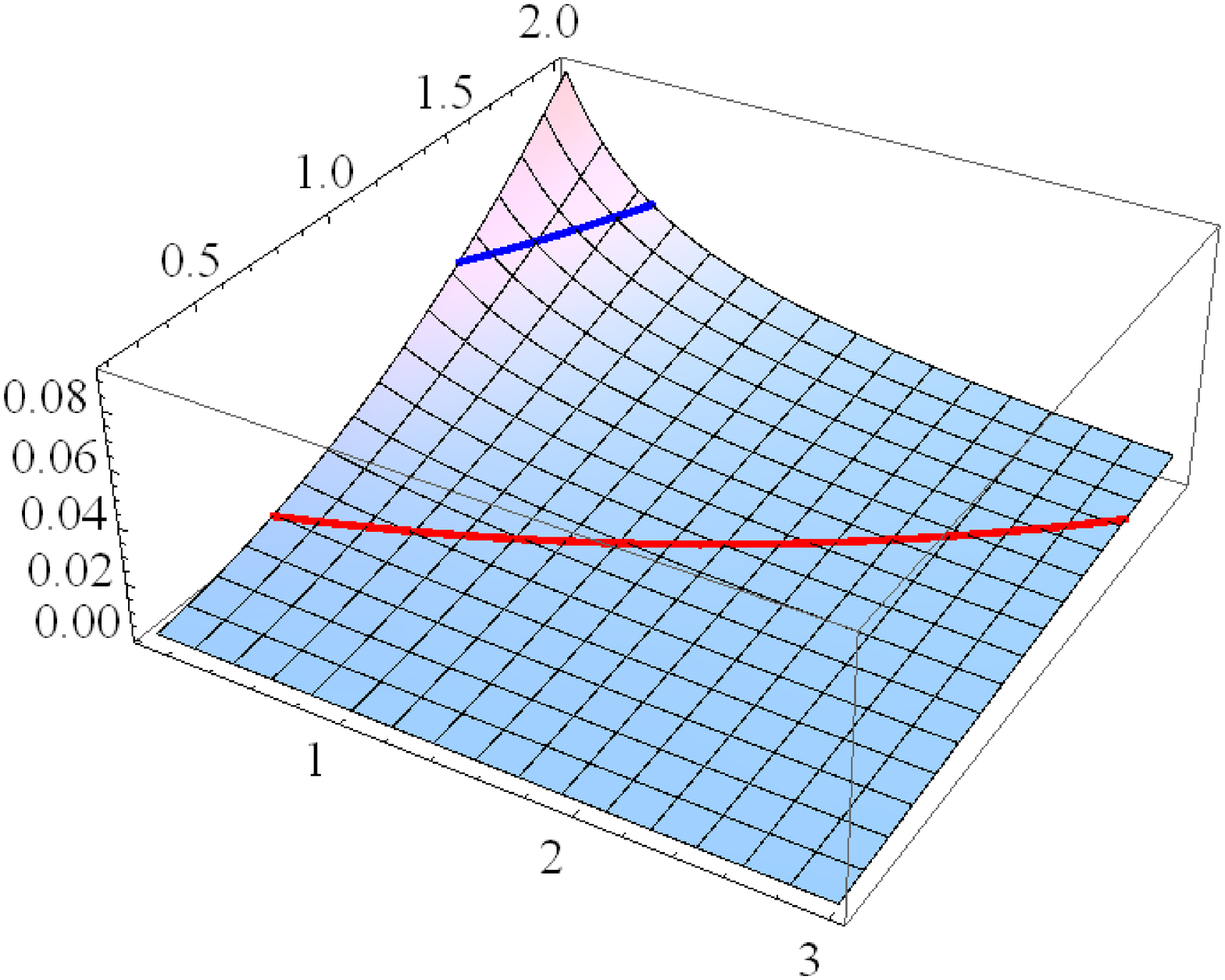}
		\put(32,10){$\lambda$}
		\put(20,68){$z_m/\rho$}
		\put(-3,27){\rotatebox{90}{$\abs{\langle\psi_0|U|\psi_0\rangle}/\omega_0$}}	
		\put(45,-3){$(b)$}	
		\put(50,28){\colorbox{white}{\small{$0.01$}}}
		\put(40,51){\colorbox{white}{\small{$0.05$}}}
	\end{overpic}
	\caption{\label{g_U}(color online)$(a)$ The ground-state expectant energy of GP on a truncated cone as a function of $\lambda,z_m/\rho$.
		$(b)$ The ratio of the ground-state expectation energy of GP to the total energy of the surface component as a function of $\lambda,z_m/\rho$.
		Contour lines with values of 0.05 and 0.01 are colored by blue and red respectively.
	}
\end{figure}

At the end of this section, we estimate the ground-state energy shift resulting from GP in a GaAs substrate. It should be made as a truncated cone-like GaAs film. Following the reference \cite{nano 24 055304}, we can obtain an out-of-plane truncated cone-like structures by etching the bulk material of the quartz substrate using an anisotropic reactive ion etch (RIE). Subsequently, we can obtain the expectant truncated cone-like GaAs film by depositing GaAs film on this quarts substrate and removing the deposition redundant. For simplicity, we consider a 2D ballistic transport model in GaAs substrate with truncated cone surface in which the representative effective mass of electron is $0.067m_e$, with $m_e$ being the mass of a rest electron.
We calculate the ground-state energy $\omega_0$ and the corresponding expectation value of the GP,  $\langle\psi_0|U|\psi_0\rangle$, as functions of $z_m$ and $\lambda$ (Fig. \ref{g_U} $(a)$) with $|\psi_0\rangle\equiv|\eta=0,\omega=\omega_0\rangle$ at $\rho=100\AA$.
The numerical results show that the absolute value of $\langle\psi_0|U|\psi_0\rangle$ is of order to be observable and increases monotonously with decrease of $z_m/\rho$ or $\lambda$.
Furthermore, increasing $\lambda$ or compressing $z_m/\rho$ will reduce the ratio of $\abs{\langle\psi_0|U|\psi_0\rangle}$ to $\omega$ monotonously.
As it shows in Fig. \ref{g_U} $(b)$, this ratio can reach nearly ten percent with $\lambda=0.1,z_m/\rho=2.0$.
In contrast, the set of $(\lambda,z_m/\rho)$ in some region would reduce this ratio to less than one percent.
Jens Gravesen and his/her coworkers have also studied the quantum dynamics for electrons constrained on a truncated cone surface model \cite{ps 72 105}.
They have presented solutions of the bound states numerically and mainly discussed the effect of thickness on surface spectra.
In this section, we give more general discussions on PDs and energy shifts relating to the geometry of the structure.

\section{Coherent electron transport in a truncated cone-like junction\label{sec_coherent}}
In this section, we will study the coherent electron transport properties on a cylindrical surface junction $S_\m{tcj}$, the surface of a truncated cone, that joins two coaxial straight cylinders with different radii.
Without loss of generality, we assume that the electron is injected from the cylinder with radius $R_1$ and is either reflected or transmitted to the cylinder with radius $R_2$.
The parametrization of $S_\m{tcj} $ is given by
\begin{eqnarray}
	\bm{r} = (\rho(z)\cos\theta,\rho(z)\sin\theta,z),
\end{eqnarray}
where $\rho(z)$ is the $z$-dependent radius of the junction and $(\theta,z)$ form the curvilinear coordinates of $S_\m{tcj}$.

\begin{figure}
	\begin{overpic}[width=0.45\textwidth,tics=5]{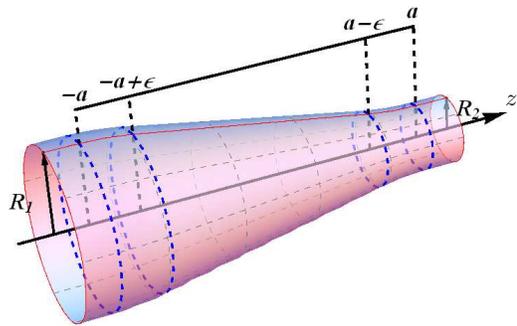}
		\put(90,45){$z$}
	\end{overpic}
	\caption{\label{f_junction} (Color online) 3D representation of a cylindrical junction that joins two straight tubes.}
\end{figure}

For realizing smooth connections between the truncated cone and the cylinders, the shape of $\rho(z)$ has been modeled with  containing parabola curves at the two ends (Fig. \ref{f_junction})
\begin{eqnarray}\label{eq_rho}
	&&\rho(z)= \n\\
	&&\left\{\begin{array}{ll}
	R_1, & z\leq -a, \\
	-\xi(z+a)^2+R_1, & -a<z\leq -a+\epsilon, \\
	-2\epsilon\xi z+ (R_1+R_2)/2, & -a+\epsilon<z\leq a-\epsilon, \\
	\xi(z-a)^2 + R_2, & a-\epsilon<z\leq a, \\
	R_2, & z> a,
	\end{array}
	\right.
\end{eqnarray}
where $\xi= (R_1-R_2)/\rb{4\epsilon a-2\epsilon^2}$ which guarantees the continuity of $\m{d}\rho/\m{d}z$, the length of junction is $2a$, and $\epsilon$ is the length of each smooth transition with $\epsilon<a$.
R. Satio and collaborators have applied the projection method to calculate the geometry of the joint of rolled-up graphene and shown that the junction joining two nanotubes with different radii can be created by rolling up the single layer graphene with connecting specific edges. \cite{prb 53 2044}.
The geometry of the nanotube proposed by R. Satio et al can be modeled with the shape of $\rho(z)$ in Eq. (\ref{eq_rho}).

The metric tensor and Weingarten curvature tensor for $S_\m{tcj} $ are determined by Eq. (\ref{eq_g_ab}) and Eq. (\ref{e_Weingarten}) respectively with replacing $f$ by $\rho(z)$, and the surface Hamiltonian $H_\m{tcj}$ with the GP $U_\m{tcj}$ on $S_\m{tcj}$ can be derived from Eq. (\ref{e_se_tangent_part2}) and Eq. (\ref{e_geo_revo}) respectively with the same substitution.
The scattering states, affected by  $U_\m{tcj}$, have been studied by numerically solving the time-independent Schr\"odinger equation $H_\m{tcj}\chi_T(\theta,z)=E \chi_T(\theta,z)$.
The quantum transmitting boundary method \cite{jap 67 6353} has been applied for open-boundary conditions on $S_\m{tcj} $.
In this structure, the open boundaries of the domain correspond to the connections between the smooth transitions on $S_\m{tcj}$ and the straight cylinders.
In the areas of $z\leq -a$ or $z\geq a$, because of the constant $\rho(z)$, the surface Hamiltonian with GP is reduced in the simple form
\begin{eqnarray}
	H_R= \frac{-\hbar^2}{2m}\rb{\frac{1}{R^2}\frac{\pd^2}{\pd\theta^2}+\frac{\pd^2}{\pd z^2}} + \frac{-\hbar^2}{2m}\frac{1}{4R^2},
\end{eqnarray}
where $R$ is $R_1\ \m{or}\ R_2$.
As the Hamiltonian on straight cylinders can be separated into longitudinal and angular components, the boundary wave functions on these structures are built as a linear combination of functions
\begin{eqnarray}\label{eq_phi_n}
	\phi_n = \sqrt{\frac{R}{2\pi}}\exp(i k_l z/\hbar)\exp(i n \theta),
\end{eqnarray}
with
\begin{eqnarray}
	\ k_l=\sqrt{2m E_l},\ R=R_1\ \m{or}\ R_2,\ \m{and}\ n= 0,1,2,\cdots.
\end{eqnarray}
In Eq. (\ref{eq_phi_n}), $\exp(ik_l z/\hbar)$ stands for a plane wave of positive longitudinal energy $E_l$, and $\sqrt{\frac{R}{2\pi}}\exp(in\theta)$ is the $n$th eigenstate of the transverse component of the cylinders.
The total energy $E$ of a particle injected in a specific transverse mode $n$ is
\begin{eqnarray}
	E = E_l+\frac{\hbar^2}{2m}\frac{n^2}{R_1^2}+U_{\m{in}},
\end{eqnarray}
where $U_\m{in}=\frac{-\hbar^2}{2m}\frac{1}{4R_1^2}$ denotes the GP in the injection cylinder.
The effective mass for an electron confined on a thin-film surface is not only determined by the particle itself, but also modified by the potential background.
In order to investigate only the curvature geometric effect on $S_\m{tcj}$, a single effective mass $m=0.173m_e$\cite{prb 72 035403}, with $m_e$ the free electron mass, has been adopted in what follows while only the transverse ground-state $n=0$ has been taken into account.

\begin{figure}
	\includegraphics[width=0.45\textwidth]{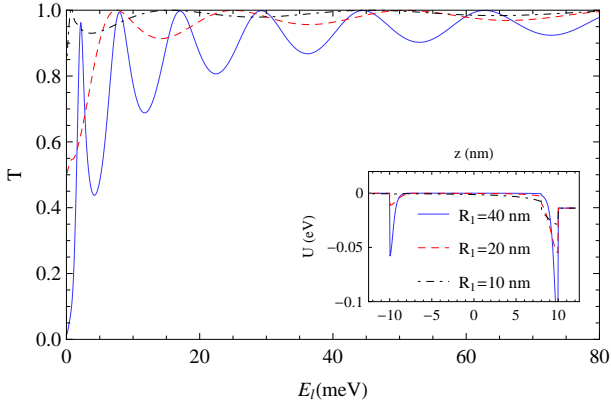}
	\caption{\label{f_TE_R1} (Color online) Transmission coefficient $T$ as a function of the injection energy for three junctions with different $R_1$. The radius of the outgoing cylinder, the length of the junction and of the smooth transitions are fixed, namely $R_2=2\m{nm}$, $a=10\m{nm}$ and $\epsilon=2\m{nm}$. $R_1=40\m{nm}$(solid curve), $R_1=20\m{nm}$(dashed curve) and $R_1=10\m{nm}$(dot-dashed curve) are taken into account. Inset: GP for three junctions described above. }
\end{figure}

\begin{figure}
	\includegraphics[width=0.45\textwidth]{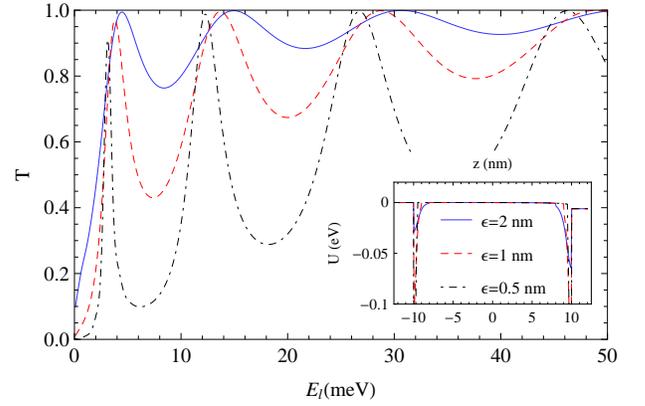}
	\caption{\label{f_TE_ep} (Color online) Transmission coefficient $T$ as a function of the injection energy $E_l$ for three junctions with different length of smooth transitions $\epsilon$. The radius of the incoming and of outgoing cylinder, and the total length of junction are fixed, namely $R_1=30\m{nm}$, $R_2=3\m{nm}$ and $2a=20\m{nm}$. Three different values $\epsilon= 2\m{nm}$ (solid curve), $\epsilon=1\m{nm}$ (dashed curve), and $\epsilon=0.5\m{nm}$ (dot-dashed curve) are considered. Inset: GP for three junctions described above. }
\end{figure}

\begin{figure}
	\includegraphics[width=0.45\textwidth]{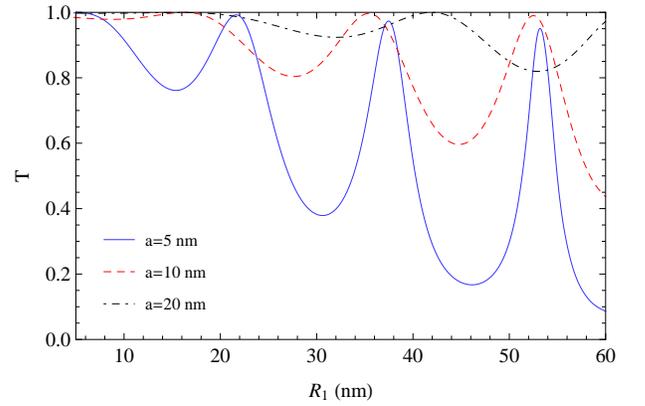}
	\caption{\label{f_T_R1} (Color online) Transmission coefficient $T$ as a function of $R_1$ for three junctions with different $a$ at $E_l=10 \m{meV}$, $R_2=2\m{nm}$ and $\epsilon=2\m{nm}$. $a=5\m{nm}$(solid curve), $a=10\m{nm}$(dashed curve) and $a=20\m{nm}$(dot-dashed curve) are taken into account.}
\end{figure}

The coherent electron transmission coefficient $T$ for $S_\m{tcj} $, as a function of the injection energy $E_l$  is reported in Fig. \ref{f_TE_R1}.
The numerical results for three different values of $R_1$, with fixed $R_2$, $\epsilon$ and $a$, has been calculated.
For a specific $R_1$, the transmission is oscillating with $E_l$ with its oscillation gradually being smoother following the increasement of $E_l$.
As the difference between $R_1$ and $R_2$ increases, the oscillations that $T$ exhibits are more pronounced.
Moreover, the average of the off-resonance transmission coefficients increases as $R_1/R_2$ decreases.
In order to study the influence on the transmission coefficient $T$ resulting from the smooth transitions, we have calculated $T$ as a function of $E_l$ for three different values of $\epsilon$ with keeping $R_1$, $R_2$ and $a$ fixed.
The results are shown in Fig. \ref{f_TE_ep}.
As $\epsilon$ decreases, the positions of the in-resonance peaks just shift a little bit to the minis direction of $E_l$, and the intervals between adjacent resonant peaks are almost unchanged.
However, oscillation in the curve with smaller $\epsilon$ exhibits greater amplitudes and more sharp resonant peaks.
Further numerical analyses show that $T$ would be nearly zero at the place between adjacent resonant peaks when $\epsilon$ is sufficiently small.
From the insets illustrating the longitudinal GP in Fig. \ref{f_TE_R1} and Fig. \ref{f_TE_ep}, it is easily seen that increasing $R_1/R_2$ and narrowing the length of smooth transitions are both available ways to form steeper GP wells at the junction ends which leads to a more pronounced oscillatory behavior of $T$ as a function of $E_l$.
However, the latter manner nearly only sharpens the GP well at the smooth transitions which mainly enlarges the amplitude of the $T-E_l$ oscillation.
Finally, for giving a better understanding of the transmission characteristics affected by the junction geometry, $T$ has been calculated as a function of $R_1$ for three junctions with different values of $a$, at a specific injection energy, $E_l=10\m{meV}$, and keeping the rest of geometric parameters fixed.
The results are reported in Fig. \ref{f_T_R1}.
It clearly shows that $T$ is functional dependent on $R_1$ with oscillation and the oscillating amplitude gradually enlarges as $R_1/R_2$ increases.
Furthermore, the resonant peaks of $T$ are less pronounced for a larger $a$.

Alex Marchi and his/her coworkers have modeled the cylindrical junction that joins two cylinders with different radii as a revolution surface with a five degree polynomial generatrix for guaranteeing a $\mathcal{C}^2$ regularity of the junction structure and a corresponding continuity of the geometric potential \cite{prb 72 035403}.
By contrast, the geometric potential (\ref{e_geo_revo}) is discontinuous because that the generatrix of the second order derivative $\rho^{\prime\prime}(z)$ is discontinuous at points with $z=-a, -a+\varepsilon, a-\varepsilon, a$.
Our model performs a better simulation of the curvature influenced transmission properties for realistic nanostructures.

\section{Conclusion\label{sec_conclusion}}
In this work, we have studied the curvature-induced bound states and the coherent transport properties for a particle constrained to move on the surface of a truncated cone.

After a short review of the thin-layer quantization scheme and the quantum dynamics on a revolution surface, we have solved the spectra on a truncated cone-like surface analytically with longitudinal hard wall boundary condition.
From the graphics of PDs in Fig. \ref{g_pd1}, it is easily seen that the non-uniform longitudinal GP induced by the surface curvature makes the constrained particles tend to distribute at the side with smaller radius.
Both the energy levels and energy differences reduce monotonously with increasing the vertex angle or $z_m/\rho$.
We have estimated the ground-state energy shift resulting from the GP, $\langle\psi_0|U|\psi_0\rangle$, for an electron strongly bound to a ballistic transport GaAs substrate with the geometry of a truncated cone.
The result shows that this expectation value is of sufficient order to be observable and  $\abs{\langle\psi_0|U|\psi_0\rangle}$ increases with reducing the $z_m/\rho$ or the vertex angle of this structure.
The ratio of $\abs{\langle\psi_0|U|\psi_0\rangle}$ to $\omega_0$ increases with reducing the vertex angle or raising the $z_m/\rho$ of the truncated cone, which demonstrates that the ratio of energy shift resulting from the GP is determined by the geometric characteristics of the structure rather than identified with the absolute value of ground-state expectation value of GP.
From the data in Fig. \ref{g_U}, it is manifest that the geometry-induced energy shift is unnegligible in some region of $(\lambda,z_m/\rho)$.

Using the quantum transmitting boundary method, we have numerically analyzed the coherent transmission coefficient for a truncated cone-like junction that joins two coaxial cylinders with different radii.
In the case of cylindrical junctions, the coherent electron transport characteristics are strongly affected by the GP .
According to the numerical results, we found that the transmission coefficient $T$ oscillates with the injection energy $E_l$.
The steep GP wells formed by geometries of the smooth transitions give a significant contribution to the resonance pattern.
The ways that steepen these GP wells, such as increasing the difference of $R_1$ and $R_2$ or decreasing the length of smooth transitions,  lead to a more pronounced oscillatory behavior of $T$ as a function of $E_l$.
In contrast, narrowing the smooth transitions strongly enlarges the amplitude of $T-E_l$ oscillation but rarely affects the values of resonant energy.
In addition, $T$ is oscillating with the geometry parameter $R_1$ when the rest of geometric parameters and injection energy are fixed.
The $T-R_1$ curves exhibit more pronounced oscillations as $R_1$ increases or the total length of junction decreases.

The model we studied is a common geometry in  nanostructures, and the methods we used are also available in studying one particle transport properties bound to thin-film structures with EM field, strain-driven geometric potential \cite{prb 84 045438} and spin-orbit interaction \cite{prb 91 245412}.

\acknowledgments
This work is supported by the National Natural Science Foundation of China (under Grant No. 11047020, No. 11404157, No. 11274166, No. 11275097, and No. 11475085), the National Basic Research Program of China (under Grant No. 2012CB921504), the Natural Science Foundation of Shandong Province of China (under Grant No. ZR2012AM022, and No. ZR2011AM019) and the Jiangsu Planned Projects for Postdoctoral Research Funds (under Grant No. 1401113C)

\end{document}